\documentclass[pra,amsmath,amssymb,superscriptaddress]{revtex4-2}

\usepackage{graphicx}% Include figure files
\usepackage{dcolumn}% Align table columns on decimal point
\usepackage{bm}% bold math
\usepackage{color}
\usepackage{appendix}

\usepackage{graphicx}
\usepackage{amsthm,amssymb}
\usepackage{array}
\usepackage{hyperref}
\usepackage{xcolor}
\usepackage{mathtools,upgreek}
\usepackage{bm,times,enumitem}
\hypersetup{colorlinks=true, urlcolor={black}, linkcolor={blue!80!black}, citecolor={blue!80!black}}

\usepackage{dsfont}
\newcommand\dsone{\mathds{1}}

\usepackage{lineno}
\begin{document}

\title{Breaking reciprocity by designed loss}

\author{I. Peshko}   

\affiliation{
B.\,I.\,Stepanov Institute of Physics, National Academy of Sciences of Belarus, Nezavisimosti Ave. 68, Minsk 220072, Belarus}

\author{D. Pustakhod}

\affiliation{Eindhoven Hendrik Casimir Institute (EHCI), Eindhoven University of Technology, 5600~MB, Eindhoven, the~Netherlands}

\author{D. Mogilevtsev}
\affiliation{ B.\,I.\,Stepanov Institute of Physics, National Academy of Sciences of Belarus, Nezavisimosti Ave. 68, Minsk 220072, Belarus}

\date{April 2022}

\begin{abstract}
In this paper we show how designed loss in open quantum systems can break reciprocity of the state propagation, and how the  non-reciprocal and even  unidirectional propagation can be achieved for different kinds of designed loss, both linear and nonlinear ones. In particular, we show how a unidirectional propagation can be achieved for states of certain symmetry in linear schemes, demonstrate possibility of building  a single-mode optical insulator by combining two kinds of nonlinear designed losses, and the way to build non-reciprocal circulator with a planar structure of dissipatively coupled waveguides. We discuss feasibility of the considered schemes and suggest possible realizations. 
\end{abstract}

\maketitle

\section{Introduction}
\label{intro}

Recently, non-reciprocal propagation of electromagnetic field became a popular research theme. Non-reciprocity is important for quite a wide of practical tasks, from field distributors and circulators to lasing and high-precision sensing \cite{pozar,Potton_2004,10.2307/1586002,PhysRevApplied.10.047001,https://doi.org/10.1002/adpr.202000104,Ma:21}. Current interest to realization of non-reciprocal systems, and especially isolators,  is born of necessity to extent methods commonly applied for longer wavelength (radio, microwaves, etc.) toward optical region and possibility to integrate non-reciprocal systems into photonic circuitry. Traditional realizations implementing magneto-sensitive media  are either hard to realize on the basis of existing integration platforms, or difficult even to realize for optical wavelength, or both \cite{PhysRevApplied.10.047001,https://doi.org/10.1002/adpr.202000104}.   So, a plethora of novel schemes for breaking reciprocity has appeared in recent years. There are schemes exploiting time-modulation \cite{soun,vanSchaijk:18}, nonlinearities \cite{shi,nader,Li:20} and interaction with few-level systems, such as atom-like structures \cite{slepyan,PhysRevApplied.10.047001,PhysRevLett.121.123601,https://doi.org/10.1002/qute.202100112},  optomechanics \cite{ewold},  topological properties of structured media \cite{https://doi.org/10.1002/adpr.202000104,RevModPhys.91.015006,lu}. 

Some quite interesting results were also obtained by considering loss.  Recently it was shown that even common single-photon  energy losses in conjunction with the usual (i.e., unitary) coupling can also be a tool for devising non-reciprocal multi-mode structures \cite{PhysRevA.82.043811,Huang2021LossinducedN}. Even more interesting results are obtained with designed loss, i.e., for example, when different systems  are coupled to the same loss reservoir inducing dissipative coupling between them, or when loss is made to be nonlinear. For instance,  designed collective  linear loss can lead to unidirectional propagation and amplification \cite{PhysRevX.5.021025,PMID:32561712}. 

Here we present another interesting feature of designed loss: it is ability to break reciprocity without conjoining with the simultaneous unitary coupling. We discuss a general recipe for doing so with the designed loss allowing for several stationary states, provide several examples and outline ways for possible practical realizations, in particular, with integrable planar waveguide systems. Notice that schemes relying only on the designed loss might be easier to realize in practice that schemes combining unitary and dissipative coupling, since one does not need adjusting different types of coupling. We show how the designed nonlinear loss can break reciprocity even for a single-mode system, and how to realize  this system using conventional Kerr nonlinearity and dissipative coupling. We also demonstrate the way to achieve non-reciprocal optical circulation using the dissipative coupling.  

The outline of the paper is as follows. Firstly, in the second Section we recall a non-reciprocity criterion for linear systems and suggest a simple generalization of the reciprocity concept suitable for  quantum state/correlation transfer, since nonlinear designed loss can produce non-classical states from the classical input. Then, in the third Section we describe the mechanics of the reciprocity breaking for the dissipatively coupled systems. In the fourth and fifth Sections we discuss examples of single-mode reciprocity breaking and unidirectional propagation with  nonlinear designed loss in asymptotic and non-asymptotic regimes, and address practical realization of such schemes.  We argue that such schemes are quite feasible  with existing photonic integration platforms and with such common modeling and simulation tools as systems of laser-written waveguides  in balk dielectric. In sixth Section we show how such common type of loss as dephasing can also break reciprocity, and demonstrate reciprocity breaking in correlations propagation for a simple device of a beam-splitter and dephaser. In seventh Section we discuss a non-reciprocal optical circulator built only with dissipatively coupled waveguides.  In the Appendix we discuss several designs of integrable dissipatively coupled structures. 

\section{Reciprocity}

First of all, let us clarify the concept of reciprocity and its breaking as applicable for the designed loss structures. We are considering here both linear and nonlinear structures, and the latter ones are known to produce non-classical 
states from classical ones  \cite{mandel_wolf_1995,PhysRevLett.83.3558,Mogilevtsev:10,PhysRevApplied.12.064051}. So, it is necessary to provide for a suitable generalization of the reciprocity concept for the open quantum systems. Here we will not pursue the task of discussing the general reciprocity concept for the open quantum systems in the manner as it was done for unitary systems, for example, in the works \cite{PhysRevA.81.032107,DEAK20121050}. Here we introduce a simple generalization of the classical definition  \cite{PhysRevApplied.10.047001,jalas,pub.1017065023,PhysRevA.81.032107,DEAK20121050}). We formulate the classical definition using quantum operators in the  following way \cite{PhysRevA.54.1661,Krasnok:19}. Let us 
define a set of $N$ input modes described by the bosonic annihilation operators  $\vec{a}=[a_1, a_2,\ldots a_N]$ (and the corresponding creation operators $a_n^{\dagger}$), and a set of $N$ output modes described by the bosonic annihilation operators  $\vec{b}=[b_1, b_2,\ldots b_N]$ (and the corresponding creation operators $b_n^{\dagger}$). For the linear device transforming input into the output, one can write \cite{PhysRevA.54.1661}:
\begin{equation}
\vec{b}=\mathbf{S} \vec{a} + \vec{N}, 
\label{qlinear}
\end{equation}
where the vector of operators $\vec{N}$ describes noise. For passive systems with $\langle\vec{N}\rangle=0$, one gets 
\begin{equation}
  \langle\vec{b}\rangle=\mathbf{S} \langle\vec{a}\rangle,
\label{classic1}
\end{equation}
where the expression $\langle x \rangle=Tr\{x\rho\}$  denotes the quantum mechanical average of the operator $x$ over the density matrix $\rho$ describing the quantum state of the whole system. For the coherent modal states, 
$\langle\vec{a}\rangle$,  $\langle\vec{b}\rangle$ are the modal amplitudes. Thus one arrives to the well-know classical description of reciprocity for the system of electromagnetic field modes.
Reciprocity is directly associated with the features of the scattering matrix $\mathbf{S}$. Non-symmetrical scattering matrix,  $\mathbf{S}\neq \mathbf{S}^T$ means that the system breaks reciprocity  \cite{PhysRevApplied.10.047001,jalas,pub.1017065023,PhysRevA.81.032107,DEAK20121050}. Notice that symmetry should be considered taking into account arbitrariness of the modal phase, i.e. possibility to make all the elements of the scattering matrix real by the local gauge transformation 
\begin{equation}
x\rightarrow x\exp\{i\phi_x\}, 
\label{loc}
\end{equation}
$x=a,b$ with some phases $\phi_x$ \cite{PhysRevA.82.043811}. 

Generally, Eq.(\ref{qlinear}) allows extending the description of reciprocity for the quantum case. For example, one can derive the transfer equation similar to Eq.(\ref{classic1}) also for the correlation function of different orders. For nonlinear transformations Eq.(\ref{qlinear}) can hardly be applied. However, one can return to linearity just by considering transformation of the density matrix elements instead of modal operators.  Let us proceed discussing reciprocity in open quantum systems in such a way as it was done for unitary systems in Refs.  \cite{PhysRevA.81.032107,DEAK20121050}.

%There are works systematically approaching quantum reciprocity for unitary dynamics \cite{PhysRevA.81.032107,DEAK20121050}, and from the thermodynamic side considering quantum generalizations of the balance equations and Onsager reciprocity relations \cite{onsager,agarwal,seifert}. 

Let us assume that our system of interest is described by the density matrix $\rho_{in}$, and we have two propagators describing propagation in different directions, $\overleftarrow{\mathcal{D}}$ and $\overrightarrow{\mathcal{D}}$. 
The state propagation corresponding to these directions is described by the mapping 
\begin{equation}
\rho_{in}\rightarrow \overrightarrow{\rho}_{out}=\overrightarrow{\mathcal{D}}\rho_{in}, \quad \rho_{in}\rightarrow \overleftarrow{\rho}_{out}=\overleftarrow{\mathcal{D}}\rho_{in}.
\label{map}
\end{equation}
Here we are not interested in the complete description of the transferred density matrix but rather in dynamics of certain observable quantities, for example, energy.  If for the same input state of our passive system the output energy for different propagation directions is different, it is a sufficient condition of the reciprocity breaking. So, let us introduce a set of Hermitian operators $P_j$ describing the observables of interest, so that quantities ${p}_j=Tr\{{P}_j\rho_{in}\}$, $\overrightarrow{p}_j=Tr\{{P}_j\overrightarrow{\rho}_{out}\}$ ($\overleftarrow{p}_j=Tr\{{P}_j\overleftarrow{\rho}_{out}\}$) can be measured at input and output of our device described by the map $\overrightarrow{\mathcal{D}}$ ($\overleftarrow{\mathcal{D}}$).   For the reciprocal transmission one would have 
\begin{equation}
\overleftarrow{p}_j=\overrightarrow{p}_j, \quad \forall j.  
\label{reci}
\end{equation}
Thus, breaking of the equality (\ref{reci}) would mean non-reciprocity. 
Notice that this simple equation implies a number of important observations. Firstly, the condition (\ref{reci}) is state dependent. So, one needs defining a class of initial density matrices, $\{\rho_{in}\}$ for discussing reciprocity. Secondly, reciprocity is defined for a particular  set  of $\{P_j\}$. This set  might be complete and sufficient for description/inference of the density matrix, or not. Different sets of $\{P_j\}$ might demonstrate both reciprocity and breaking of it for the same system. Only proving reciprocity for arbitrary initial states and observables  allows calling the system completely reciprocal. 

\begin{figure}[htb]
\begin{center}
\includegraphics[width=\linewidth]{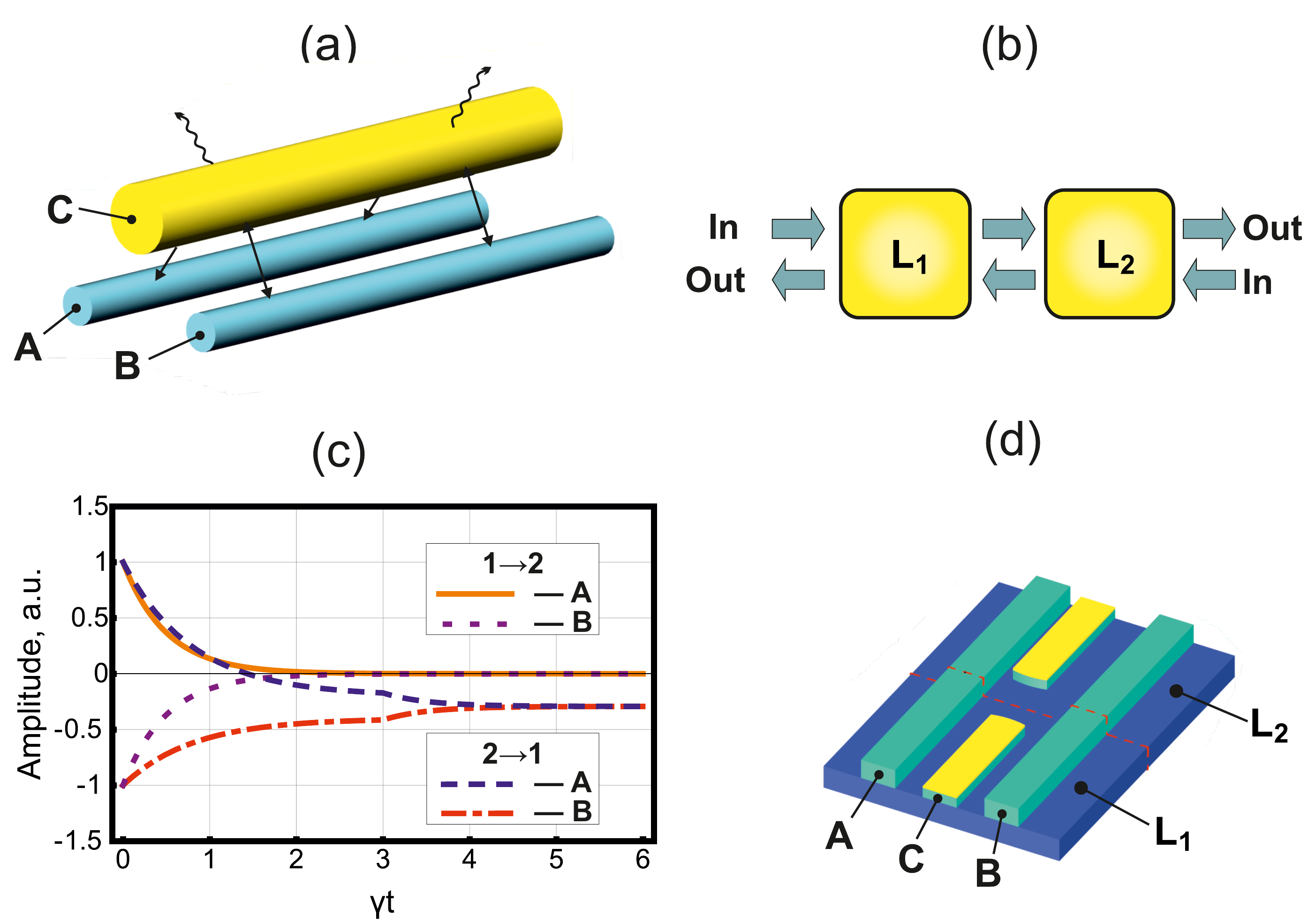}%{fig1.pdf}
\end{center}
\caption{(a) A scheme of the dissipative coupling of two waveguides A and B through the common loss reservoir represented as a lossy waveguide C. (b) An illustration of the modal amplitude dynamics for the direct and reverse propagation through the two waveguides with different dissipative coupling in the first part (denoted as $L_1$ in the panel (c)) and in the second part (denoted as $L_2$ in the panel (c)). Solid and dotted curves correspond to the field propagation in the waveguides A and B in the direction $1\rightarrow 2$, the dashed and dash-dotted lines correspond to the field propagation in the waveguides A and B in the direction $2\rightarrow 1$. For the first device part the Lindblad operator is $L_1=\gamma (a-0.5b)$, for the second part the Lindblad operator is $L_2=\gamma(a-b)$. The initial amplitude for the mode $a$ is 1, for the mode $b$ is -1.  Panel (d) depicts a scheme of the integrable device combining two different kinds of  dissipative coupling by variable placing of a central lossy waveguide between two low-loss waveguides (see discussion in Appendix).}
\label{fig1}
\end{figure}

\section{Breaking reciprocity by linear dissipative coupling}

First of all, let us demonstrate how a purely dissipative device breaks reciprocity. We consider a set of dissipatively coupled waveguides without an additional unitary coupling between them. Such a coupling can be realized by usual unitary coupling of several waveguides to the common dissipative reservoir, say, the waveguide with strong loss (as depicted in Fig.~\ref{fig1}(a)). Such an arrangement can be easily realized in practice, for example, by laser-writing a systems of single-mode waveguides in a balk glass \cite{bigger,Mogilevtsev:10,Eichelkraut:14,PhysRevApplied.12.064051,Mukherjee2017DissipativelyCW}, or growing planar semiconductor structures on  some platform, for example, InGaAsP material on  InP  \cite{Ma:21,jstqe-augu18} (see the Appendix for discussion of several designs of dissipative waveguide couplers on the integrated photonic platform).  

To demonstrate a breaking of reciprocity, let us consider a bipartite two-waveguides system as schematically shown in Fig.~\ref{fig1}(c). During propagation through the $j$-th part of the system, the density matrix $\rho$ dynamics is described by the following master equation 
\begin{equation}
\frac{d}{dt}\rho=\gamma_j\left(2L_j\rho L_j^{\dagger}-\rho L_j^{\dagger}L_j-L_j^{\dagger}L_j\rho\right),
\label{lind}
\end{equation}
where the Lindblad operators for the $j$-th part are taken as $L_j=a+\delta_jb$, $a$ and $b$ being the annihilation operators for the modes A and B, $\delta_j$ are asymmetry parameters describing the respective strength of waveguides coupling to the common reservoir;  $\gamma_j$ are decay rates corresponding to each part.  Taking into account a possibility of the local gauge transformation (\ref{loc}), one can assume real $\delta_j$ (this we also assume in the further considerations). 
In Fig.~\ref{fig1}(b) one can see an illustration of the reciprocity breaking for the bipartite systems with $L_1=\gamma (a-0.5b)$ and  $L_2=\gamma(a-b)$. For the asymmetric input state of both modes (i.e. when the sum of the amplitudes is zero),  one  has a unidirectional energy flow through the system. For such an input,  the action of the second part of the device just exponentially reduces the state to the vacuum. 

Such behavior can be easily understood from the character of dynamics dictated by the master equation (\ref{lind}).  The dissipative coupling between modes in $j$-th part of the system  drives their state to the one satisfying $L_j\rho=0$ \cite{Mogilevtsev:10,Mukherjee2017DissipativelyCW}. Asymptotically (i.e. for the interaction time tending to infinity), the modal amplitude transformation can be described by the following scattering matrix
\begin{equation}
\begin{bmatrix} \langle a_{out} \rangle \\ \langle b_{out} \rangle \end{bmatrix} = \mathbf{S}_j \begin{bmatrix} \langle a_{in} \rangle \\ \langle b_{in} \rangle \end{bmatrix}, \quad \mathbf{S}_j=\frac{1}{1+\delta_j^2} \begin{bmatrix} \delta_j^2 & -\delta_j \\ -\delta_j & 1 \end{bmatrix}. 
\label{sdis}
\end{equation}
Generally, for $\delta_1\neq \delta_2$ and $\delta_1\delta_2\neq -1$, the product of the matrices  $\mathbf{S}_1$ and $\mathbf{S}_2$ is not symmetric
\begin{equation}
\mathbf{S}_1\mathbf{S}_2= \frac{1+\delta_1\delta_2}{(1+\delta_1^2)(1+\delta_2^2)}\begin{bmatrix} \delta_1\delta_2 & -\delta_1 \\ -\delta_2 & 1 \end{bmatrix}. 
\label{prodmat}
\end{equation}
So, our bipartite device is indeed not reciprocal. Moreover, it does not allows the states satisfying $\delta_1\langle a_{in} \rangle-\langle b_{in} \rangle=0$  to propagate in the direction $1 \rightarrow 2$, and it does not allows the states satisfying $\delta_2\langle a_{in} \rangle-\langle b_{in}\rangle=0$ to propagate in the opposite direction. Notice that different dissipative coupling in the realization schematically depicted in Fig.~\ref{fig1}(a) can be simply realized by adjusting coupling with the common reservoirs, for example, by adjusting distances between the  waveguides A and B and the central dissipative waveguide as shown in Fig.~\ref{fig1}(d) \cite{PhysRevApplied.12.064051,Mukherjee2017DissipativelyCW}. 

Interestingly, that a simple linear device discussed here can offer quite considerable asymmetry in energy exchange between waveguides  for different propagation directions. Indeed, in is easy to get from Eq.(\ref{prodmat}) that the ratio of the off-diagonal elements of the matrix products 
\begin{equation}
\left|\frac{[\mathbf{S}_1\mathbf{S}_2]_{12}}{[\mathbf{S}_2\mathbf{S}_1]_{12}} \right|=\left|\frac{\delta_1}{\delta_2}\right|
\label{ratio1}
\end{equation}
is proportional to the ratio of the asymmetry parameters. As long as the value of $\delta_j\gamma$ remains much larger than the rate of uncorrelated  single-photon loss inevitably present in a realistic waveguide system, one can get $\left|\frac{\delta_1}{\delta_2}\right|^2$ difference in energy transmission between waveguides in different directions. For example, for planar systems described in Appendix it seems realistic to reach the ratio of 0.1 between waveguide coupling rates in the scheme depicted in Fig.~\ref{fig1}(d), thus to have $\left|\frac{\delta_1}{\delta_2}\right|^2=0.01$ and 20 dB in energy exchange asymmetry.

Naturally, our discussion can be easily extended for more involved dissipatively coupled system of modes. The main message would be the same: conjunction of parts with different dissipative coupling leads to breaking of reciprocity. 

Also, it is useful to note that breaking reciprocity can be achieved with only one of two parts being a dissipative coupler. Indeed, replacing, for example, the first part with the beam-splitter transforming some set of input states to the ones satisfying $\delta_2\langle a_{in} \rangle-\langle b_{in}\rangle=0$, one gets unidirectional propagation of energy for this set of states (only for  $2 \rightarrow 1$). Also, to break reciprocity one can simply introduce different  uncorrelated loss in both modes of the first part. Indeed, for the uncorrelated loss in both waveguides $A$ and $B$ (taken to be finite-length) the scattering matrix is $\mathbf{S}_1=\mathrm{diag}\{T_A,T_B\}$, where $T_A$, $T_B$ are  transmission coefficients for the modes $A$ and $B$. We assume the asymptotic regime for the second, dissipative coupler part. So, for the transmission through the bipartite compound device one has the following scattering matrix 
\begin{equation}
\mathbf{S}_2\mathbf{S}_1= \frac{1}{1+\delta_2^2}\begin{bmatrix} T_A\delta_2 & -T_B\delta_2 \\ -\delta_2T_A & T_B \end{bmatrix}. 
\label{prodmat1}
\end{equation}
The product (\ref{prodmat1}) is asymmetric. The asymmetry in energy exchange between waveguides then is defined by the ratio $(T_A/T_B)^2$.

\begin{figure}[htb]
\begin{center}
\includegraphics[width=0.75\linewidth]{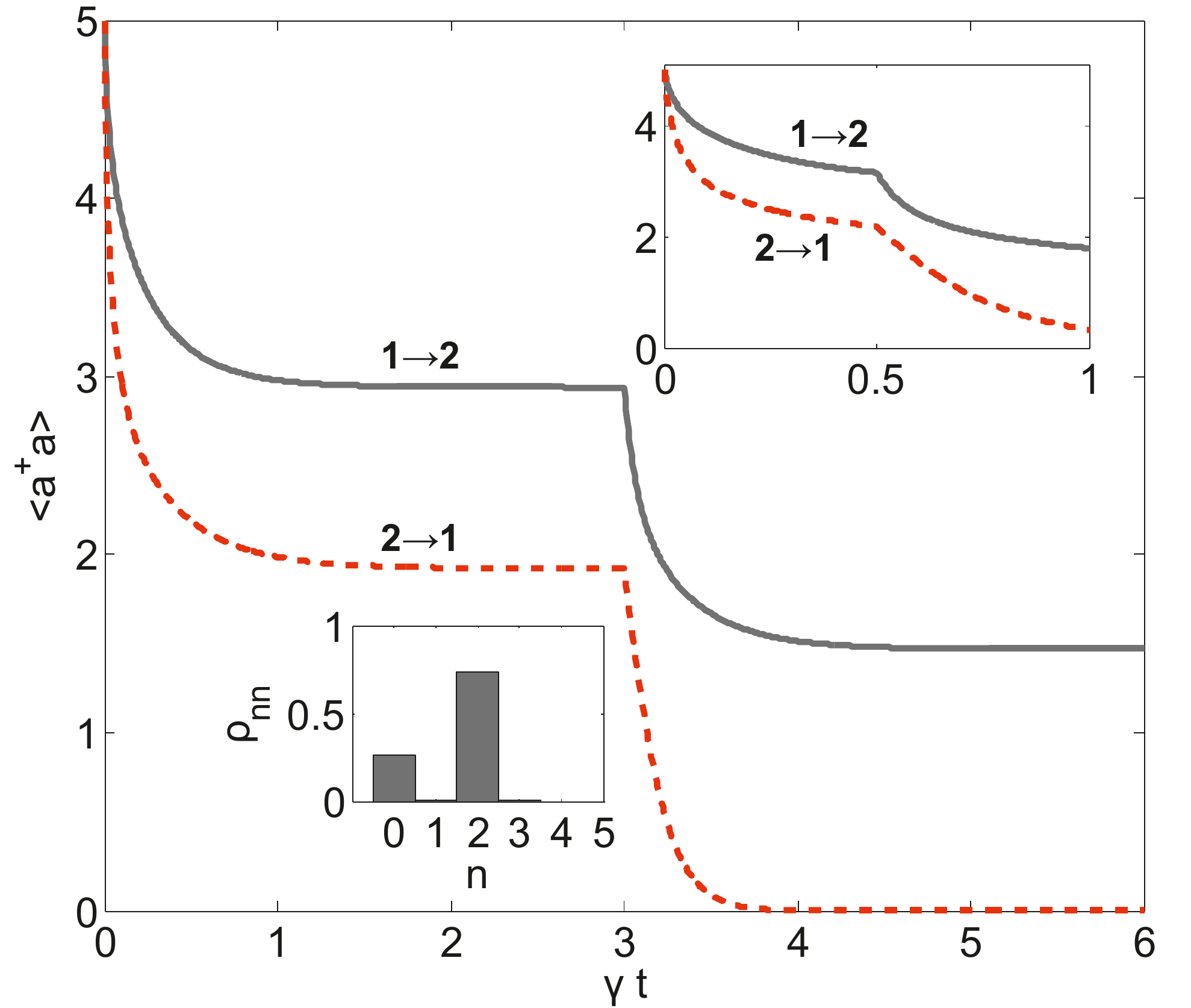}%{fig1.pdf}
\end{center}
\caption{Dynamics of the population for the single-mode bipartite system with parts described by the Lindblad operators (\ref{lm}) with $m_1=4$ and $m_2=2$ as given by the master equation (\ref{lind}) for the initial coherent state with 5 photons and the same rate $\gamma$. Solid lines both in the main panel and in the upper inset correspond to the propagation $1\rightarrow 2$; dashed lines both in the main panel and the inset correspond to the propagation $2\rightarrow 1$. For the main panel propagation through each part occurs during the interval  $3\gamma t$; for the upper panel propagation through each part occurs during the interval  $0.5\gamma t$. Lower panel shows the photon number distribution of the output state for the propagation $1\rightarrow 2$ depicted in the main panel.}
\label{fig2}
\end{figure}

\section{Breaking reciprocity with nonlinear dissipation}

The obvious idea of breaking commutativity of dynamics in different parts of a compound system to break reciprocity can be easily extended for the general dissipative dynamics. Generally, the state transformation of the open system  can be described  in a standard way with help of the Kraus operators $D_j$ as 
\begin{equation}
\rho_{out}=\sum\limits_{\forall j}D_j\rho_{in} D^{\dagger}_j, 
\label{kraus}
\end{equation}
where the Kraus operators satisfy 
$\sum\limits_{\forall j}D_j^{\dagger} D_j=\dsone$ \cite{kraus}. 

It is easy to surmise that for our bipartite system as depicted in Fig.~\ref{fig1}(c), non-commutativity of the Kraus  operators $D_j^{(1)}$ and $D_k^{(2)}$ corresponding to the first and the second parts of the system, can also lead to breaking of reciprocity.  Now let us show that the dissipative systems possessing several stationary states (similarly to the linear case described above) can also break reciprocity. Moreover, in stark difference with the linear case, such system can provide for a single-mode non-reciprocity. 

Indeed, let us consider a simple  example of single-mode subsystems 1 and 2 with  just two stationary states each. These systems are described by the following Kraus operators
\begin{eqnarray}
D_{m}^{(1,2)} =
    \begin{cases}
      |m_{1,2}\rangle\langle m| & \forall m\geq m_{1,2}\\
      |0\rangle\langle m| & \forall m < m_{1,2},
    \end{cases}       
\label{kraus2}
\end{eqnarray}
where the states $|m\rangle$ are Fock states with $m$ photons of the single considered mode, and $m_{1,2}\neq 0$. For $m_1\neq m_2$ some Kraus operators corresponding to different parts are obviously not commuting, so, the state propagation is not reciprocal. Moreover, combination of the subsystems 1 and 2 realizes a single-mode optical insulator. Indeed, for $m_1>m_2$, any Fock-state input with $m>m_1$ produces Fock-state with $m_2$ photons when propagating $1\rightarrow 2$. Whereas any state input from the opposite side (i.e., when $2\rightarrow 1$) produces just the vacuum state.  

Despite its seemingly idealized character, the scheme with 
the Kraus operators (\ref{kraus2}) can be realized, for example, with nonlinear coherent loss described by the Lindblad operators 
\begin{equation}
 L_j=a(a^{\dagger}a-m_j), \quad j=1,2,   
 \label{lm}
\end{equation} 
$m$ being a positive integer number \cite{Mikhalychev_2011}. 

It is interesting that the regime of optical isolation with the device described by the Lindblad operators (\ref{lm}) functions also with classical inputs. An example of such functioning is shown in Fig.~\ref{fig2}. The main panel of Fig.~\ref{fig2} shows how the average number of photons behaves for the input at the part with $m_1=4$ (solid line) and output at the  part with $m_2=2$. Dashed lines shows dynamics of the average number of photons for propagation in the opposite direction. Low inset shows the output state for the propagation $1\rightarrow 2$ is non-classical being a mixture of the vacuum with the two-photon Fock state. 

Another interesting feature with the scheme  (\ref{lm}) is that non-reciprocity can be quite pronounced even far from asymptotic regime, i.e., when the state  of the mode is far form the one given by the condition $L_j\rho=0$. This situation is illustrated in the upper panel in Fig.~\ref{fig2}. Actually, this feature makes it feasible looking for practical realization of non-reciprocity with designed nonlinear loss of the considered kind. Indeed, close to the stationary state satisfying $L_j\rho=0$, the dynamics resulted from non-linear loss is rather slow and can be easily disrupted by common single-photon loss unavoidable in realistic systems \cite{Mogilevtsev:10,PhysRevApplied.12.064051}.

\begin{figure}[htb]
\begin{center}
\includegraphics[width=0.55\linewidth]{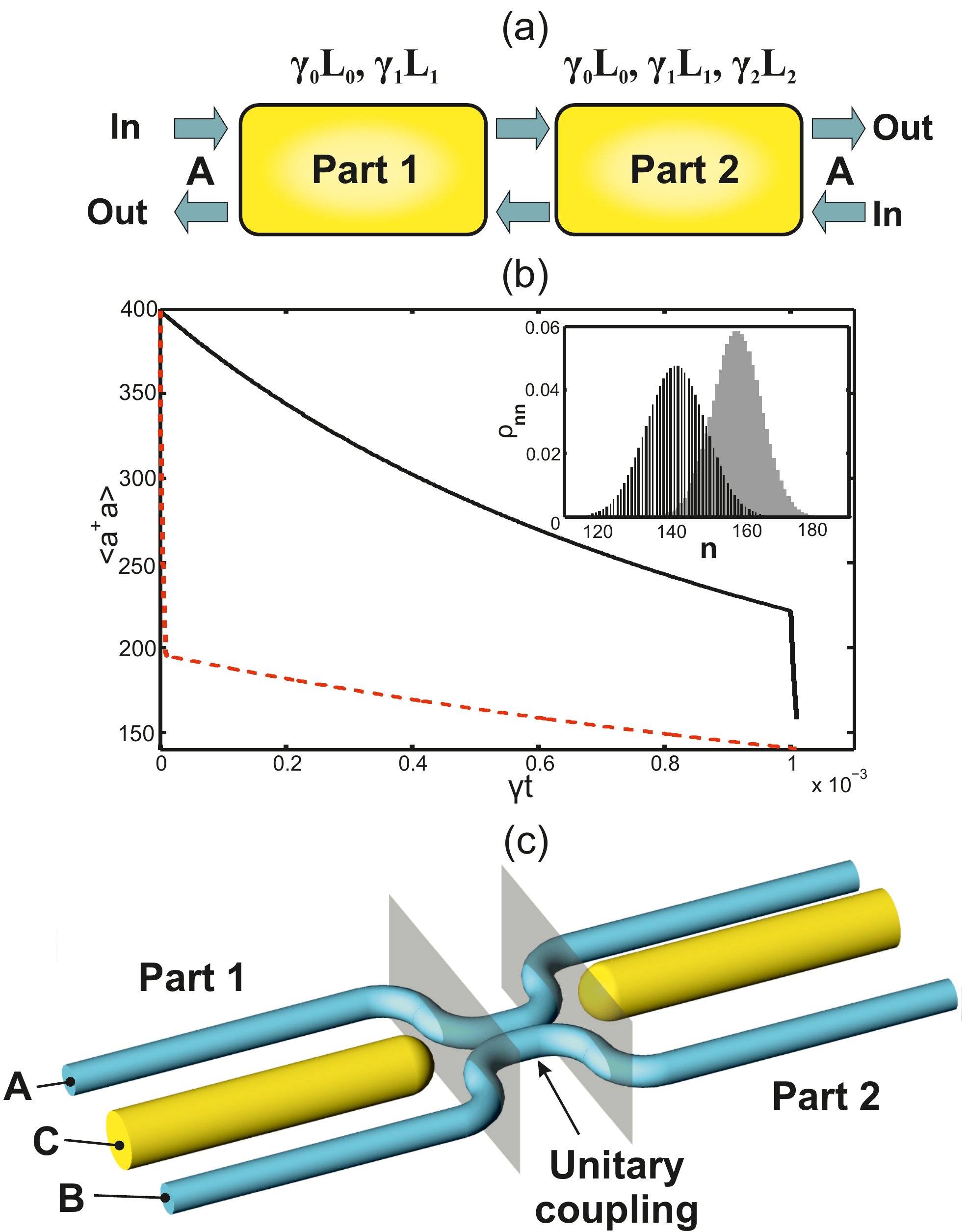}
\end{center}
\caption{(a) The scheme of the single-mode bipartite non-reciprocal device described by Eq. (\ref{lind2}). 
The panel (b) shows dynamics of the average number of photons for the structure schematically depicted in the panel (a). The solid line corresponds to propagation $1 \rightarrow 2$, the dashed line corresponds to the propagation in the opposite direction. For the Part 1 the dynamics is described by the master equation (\ref{lind2}) with decay rates $\gamma_0=\gamma_1=\gamma$, $\gamma_2=0$.  For the Part 2 the decay rates are $\gamma_{0,1,2}=\gamma$. The unitary transformation between the two parts is supposed to leave the collective mode state unchanged. The initial state at each input is taken as the coherent state of the corresponding "stationary" collective mode  with the amplitude $\alpha=20$. The inset shows the output state  photon number distribution, $\rho_{nn}$, for the propagation $1\rightarrow 2$ (gray thick bars) and for the propagation in the opposite direction, $2\rightarrow 1$ (thin black bars). (c) An example of the bipartite nonlinear loss scheme for realizing non-asymptotic breaking of reciprocity by single-mode nonlinear loss.  Two self-Kerr nonlinear waveguides are coupled to the third waveguide, but not to each other. The third waveguide is designed to have excess loss. In the part 1 the third waveguide is coupled symmetrically to the waveguides A and B realizing the two-photon loss. In the part 2 the coupling is asymmetric realizing both the two-photon loss and the nonlinear coherent loss (NCL). We assume a unitary coupling  between the parts 1 and 2 allowing for lossless transformation of  a collective mode $\tilde{A_1}$ to the collective mode $\tilde{A_2}$ and vice versa.  }
\label{fig3}
\end{figure}

\section{Toward practical non-reciprocity with nonlinear dissipation}

As it was already mentioned here, practical realization of the dissipative coupling between waveguides is rather simple to achieve, for example, by laser-writing single-mode waveguides in glass \cite{bigger,Mogilevtsev:10,Eichelkraut:14,PhysRevApplied.12.064051,Mukherjee2017DissipativelyCW}, or fabricating planar structures as described in the Appendix  (and one might mention that waveguide-writing in borosilicate glass has become a very versatile photonic instrument for modeling a plethora of exotic physical effects (see, for example, Refs.\cite{Mukherjee2017DissipativelyCW,PhysRevLett.121.075502})). Designing nonlinear loss is more difficult and involved. Some schemes were suggested and experimentally demonstrated with trapped ions and atoms, nanomechanical resonators, superconducting microwave circuits, nonlinear Kerr resonators, etc.  \cite{PhysRevLett.77.4728,PhysRevLett.86.4988,2009NatPh...5..633V,RevModPhys.75.281,PMID:30728523,PhysRevB.70.205304}. To demonstrate the way of experimental realization of reciprocity breaking purely by designed nonlinear loss, here we consider a recently suggested scheme with self-Kerr nonlinear waveguides that implements the dissipative coupling scheme depicted in Fig.~\ref{fig1}(a) \cite{Mogilevtsev:10,PhysRevApplied.12.064051,2020}. This scheme allows to produce two-photon loss or nonlinear coherent loss of the kind given  by Eq.(\ref{lm}) for $m_j=1$ combined with two-photon loss for the essentially single-mode device.  

Dynamics of the single-mode system considered in Refs. \cite{Mogilevtsev:10,PhysRevApplied.12.064051} is described by the following master equation
\begin{eqnarray}
\frac{d}{dt}\rho=\sum\limits_{j=0}^2\gamma_j\Bigl(2L_j\rho L_j^{\dagger}-
\rho L_j^{\dagger}L_j-L_j^{\dagger}L_j\rho\Bigr),
\label{lind2}
\end{eqnarray}
where $L_0=a$ describes common single-photon loss; $L_1=a^2$ describes two-photon loss and $L_2=a(a^{\dagger}a-1)$ describes so called nonlinear coherent loss (NCL) \cite{Mikhalychev_2011}; $\gamma_{0,1,2}$ are the corresponding loss rates. 

Firstly, let us show that the bipartite system composed of the parts described by the master equation (\ref{lind2}) with different sets of decay rates $\gamma_j$ can lead to asymptotic breaking of reciprocity. To that end we assume that for the first part we have $\gamma_{0,2}=0$, $\gamma_1=\gamma$ and for the second part we have $\gamma_{0,1}=0$, $\gamma_2=\gamma$ in Eq.(\ref{lind2}). I.e., we have only  two-photon loss in the first part and only the NCL in the second part of our device.  Two-photon loss does not affect states in the subspace of single and zero photon states. Asymptotically, two-photon loss  drives an initial  state with the number of photon much larger than unity toward a mixed single-photon state with one-half photons on average \cite{PhysRevLett.85.1136,PhysRevLett.83.3558}. The NCL described by $L_2$ also does not affect states in the subspace of single and zero photon states. But asymptotically it drives an initial  state with the number of photon much larger than unity toward the single-photon state \cite{Mogilevtsev:10,Mikhalychev_2011}. So, for the initial  coherent state with the average number of photons much larger than unity, the bipartite system will produce asymptotically the state with $1/2$ photon on average for the $1\rightarrow 2$ propagation, and the single-photon state for propagation in the opposite direction.  

Fig.~\ref{fig3}(b) shows that reciprocity is broken for such a system also for a finite 
time even in the regime when the systems' state is far from the stationary state.  For the simulation we have taken that the system is also subject to the common single-photon loss, and that the two-photon loss is present also in the second part of the device (which corresponds to the realistic three-waveguide scheme \cite{PhysRevApplied.12.064051,2020}). So, for the first part we have taken $\gamma_0=\gamma_1=\gamma$, $\gamma_2=0$; for the second part we have taken $\gamma_j=\gamma$, $\forall j$.  The initial state is taken to be the coherent one with the amplitude $20$. One can see from Fig.~\ref{fig3}(b) that the average number of photons at outputs is different, and the photon number distributions of the output states (shown in the inset in (b)) are also quite different. Both output states are sub-Poissonian, but for  the $1\rightarrow 2$ propagation the state is considerable more photon-number squeezed, with the Mandel parameter $Q\approx -0.7$. Whereas in the opposite direction, $Q\approx -0.5$. 

The single-mode device analyzed above can be realized with the waveguide arrangement schematically shown in Fig.~\ref{fig3}(c). The device consists of three parts: the part 1 and part 2 realize different kinds of nonlinear loss by the dissipative coupling scheme shown in Fig.~\ref{fig1}(a). The symmetric coupling between the side self-Kerr nonlinear waveguides   and the lossy central one allows to realize the two-photon loss of the   collective mode $A_1\propto a_1-b_1$, where $a_{1}$, $b_{1}$ are the modes guided in corresponding waveguides. The mode $A_1$ would be a stationary mode of the part 1 in absence of nonlinearity and uncorrelated single-photon loss. In the part 2 the coupling is asymmetric, so, NCL arises described by the Lindblad operator $L_2$ in addition to the two-photon loss for the collective mode $A_2\propto \delta a_2 -b_2$, the parameter $\delta$ describing the coupling asymmetry \cite{Mogilevtsev:10,Mikhalychev_2011}. Notice that for initial states with large number of photons, on the initial stage of dynamics (i.e., when the average number of photons still remains large) NCL dominates over two-photon loss making its influence on the photon number dynamics negligible even though  the loss rates are close in value \cite{PhysRevApplied.12.064051}). The third part connecting  parts 1 and 2 in Fig.~\ref{fig3}(a) provides for the effectively single-mode operation of the device transforming the collective mode $A_1$ into $A_2$ and vice versa. Initially  one should also excite only $A_1$ for the propagation $1\rightarrow 2$, and $A_2$ from the propagation in the opposite direction. 

Notice that for demonstrating non-reciprocity one might simplify the set-up of Fig.~\ref{fig3}(c) getting rid of the central unitary coupled part, i.e., leaving only two conjoined nonlinear dissipative couplers as in Fig.~\ref{fig1}(d) . But the single-modality of the operation would be lost, and both linear (considered in Section 3) and nonlinear effects of the reciprocity breaking would be combined. 

 \section{Breaking reciprocity by dephasing}
 
 We have already shown with examples of both linear and nonlinear  dissipative dynamics that dissipative systems possessing several stationary states can break reciprocity. Now let us consider another example of quite simple and widespread (actually, even virtually ubiquitous) kind of dissipative dynamics with several (and possibly even infinite number of) stationary states that can also break reciprocity. Also, this reciprocity violation cannot be captured by a simple classical criterion implementing the amplitude scattering matrix  (\ref{sdis}). 
 
 This kind of dissipation is dephasing \cite{breuer2002theory}. The essence of it is as follows. The dephasing with respect to some orthogonal basis $\{|\psi_j\rangle\}$ (i.e, $\langle\psi_k|\psi_j\rangle=\delta_{kj}$) means that asymptotically any state is transformed to the state with only the diagonal components in this basis
 \begin{equation}
 \rho=\sum\limits_{\forall j,k}  c_{jk}|\psi_j\rangle\langle\psi_k| \rightarrow \bar{\rho}=\sum\limits_{\forall j}  c_{jj}|\psi_j\rangle\langle\psi_j|, 
     \label{dephas}
 \end{equation}
where the coefficients $c_{jk}$ describe the representation of the state density matrix in the basis   $\{|\psi_j\rangle\}$. The transformation (\ref{dephas}) is realized by the Kraus operators $D_{j}=|\psi_j\rangle\langle\psi_j|$, and any state commuting with all these Kraus operators is stationary.   

It is easy to see that a bipartite system with each part performing dephasing  with respect to different basis breaks reciprocity. Indeed, let us take that the $j$-th part  dephases the input state with  respect to the basis $\{|\psi_k^{(j)}\rangle\}$, and at least for  some $j$ and $k$ one has $\langle\psi_k^{(1)}|\psi_j^{(2)}\rangle\neq 0$. Than the  corresponding Kraus operators $D^{(1)_k}$ and $D^{(2)_j}$ do not commute and the dynamics is not reciprocal.

Physically, realization of a single mode dephasing in different bases is not simple and involves state transformation corresponding to transition to the different basis. However, reciprocity breaking can be realized conjoining unitary and feasible dephasing parts. Dephasing in the energy basis is quite simply realized by random fluctuation of frequency, or, equivalently, by inducing random fluctuation of modal phase \cite{breuer2002theory}. Let us show how it is possible to break reciprocity in propagation of correlations by a bipartite system consisting of a simple 50/50 beam-splitter and the energy dephaser.   

We take a two-mode system as shown in Fig.~\ref{fig1}(c). The part 1 is a beam-splitter performing the following standard transformation 
\begin{equation}
    \label{bs}   a_{out}=\frac{1}{\sqrt{2}}(a_{in}+ib_{in}), \quad b_{out}=\frac{1}{\sqrt{2}}(ia_{in}+b_{in}).
\end{equation}
The part 2 is the dephaser in the energy basis. Let us consider the following pure input state
\begin{equation}
\rho_{in}=|\psi\rangle\langle\psi|, \quad |\psi\rangle=x|1_a\rangle|0_b\rangle+\sqrt{1-x^2}|0_a\rangle|1_b\rangle,
    \label{inst}   
\end{equation}
where the vector $|n_x\rangle$ is the Fock state of the mode $x$ with $n_x$ photons, and a real $x$ satisfies $|x|\leq 1$. 

Now as it was discussed in Section 2, let us consider an observable $P=\lambda a^{\dagger}b+\lambda^*b^{\dagger}a$, where $\lambda$ is a scalar parameter.  It is easy to see that the value of this observable for the initial state (\ref{inst}) is 
\[ p=\mathrm{Tr}\{P\rho_{in}\}=(\lambda+\lambda^*)x\sqrt{1-x^2}.
\]
For the propagation $1\rightarrow 2$ we have from Eqs.(\ref{bs},\ref{inst}) $\overrightarrow{p}=0$. For the propagation $2\rightarrow 1$ we have 
\[\overleftarrow{p}=\frac{i}{2}(\lambda-\lambda^*)(1-2x^2). 
\]
So, according to the definition given by Eqs.(\ref{map},\ref{reci}), for $\lambda\neq\lambda^*$ and $|x|\neq 1/\sqrt{2}$ our device is indeed breaking reciprocity. In our case, modal correlations are non-reciprocally transmitted through our bipartite device.  Notice that the classical definition (\ref{classic1}) cannot work for the case. For the initial state (\ref{inst}) modal amplitudes are zero.

\begin{figure}[htb]
\begin{center}
\includegraphics[width=0.8\linewidth]{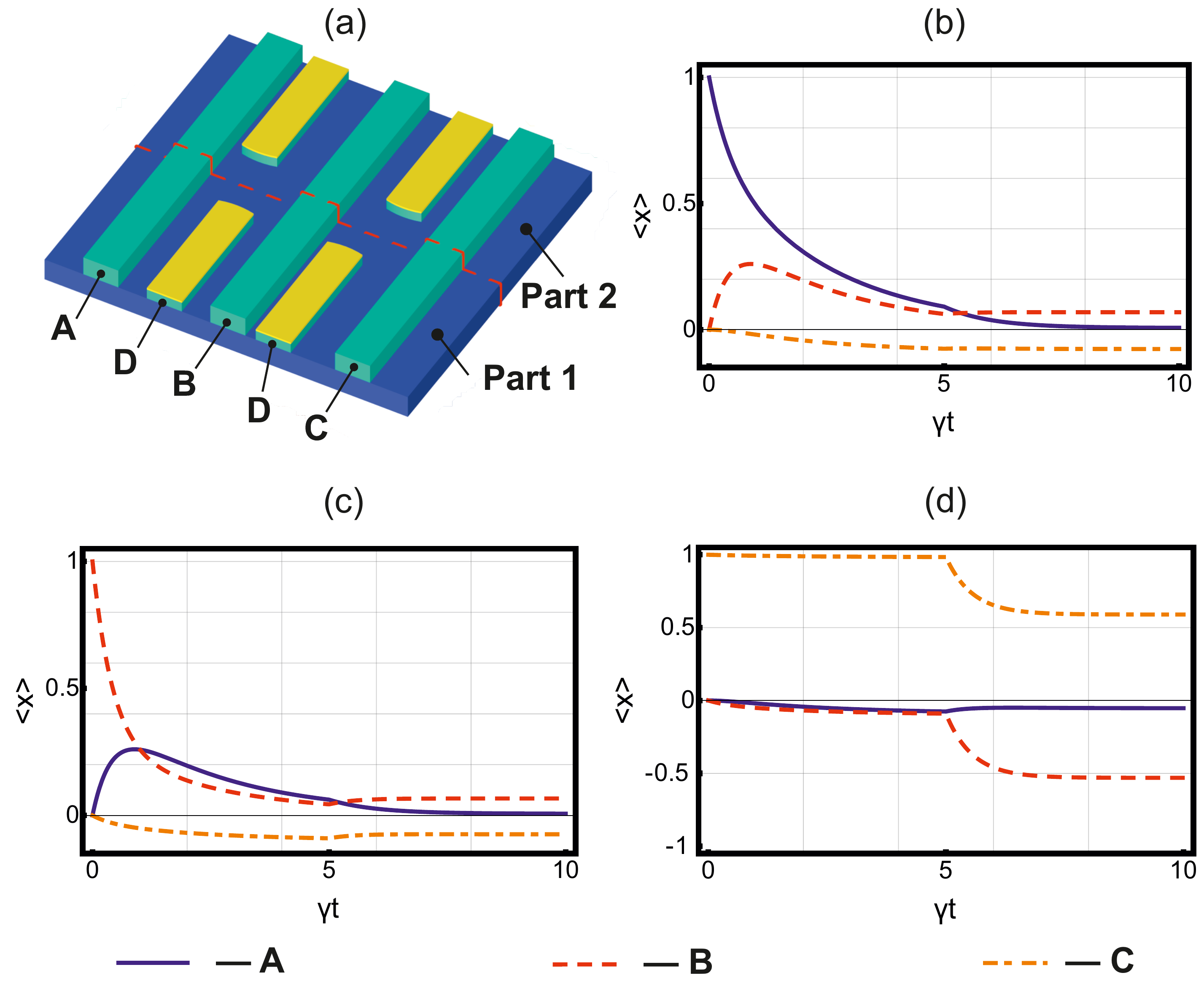}
\end{center}
\caption{(a) A scheme of bipartite three-mode dissipatively coupled rotator. We consider only the propagation $1\rightarrow 2$. Panels (b-d) shows dynamics of modal amplitudes as given by Eqs. (\ref{lind3},\ref{ltr}) with decay rates $\gamma_{x,j}=\gamma$, $\forall x,j $, and asymmetry parameters $y_1=-z_2=-0.9$, $y_2=-z_1=0.1$. The border between the parts corresponds to $\gamma t=5$. In the panels (b-d) solid, dashed and dash-dotted lines correspond to the modal amplitudes of the modes in the waveguides $x = A, B, C$. The panels  (b,c,d) correspond to the initial coherent excitation with the unit amplitude of the modes of the waveguides A,B,C with vacuum of other two modes.} 
\label{fig4}
\end{figure}

\section{Asymmetric dissipatively coupled circulator}

Finally, let us demonstrate how purely dissipative coupling might serve for building an asymmetric optical "circulator",  i.e. a device that asymmetrically transfers modal energy between the modes in dependence on the initial excitation. We consider a  bipartite structure with each part having three physical modes A, B, C,  where dissipative coupling is only between the modes A,B and B,C (which makes the structure feasible to fabricate using planar arrangement of strongly coupled low-loss waveguides A, B, C and high-loss waveguides D as shown in Fig.~\ref{fig4}(a); such a structure is a simple generalization of ones discussed in the Appendix).  

Each part of our bipartite set-up is described by the following master equation ($j=1,2$)
\begin{eqnarray}
\frac{d}{dt}\rho=\sum\limits_{x=AB,BC}\gamma_{x,j}\Bigl(2L_{x,j}\rho L_{x,j}^{\dagger}-
\rho L_{x,j}^{\dagger}L_{x,j}-L_{x,j}^{\dagger}L_{x,j}\rho\Bigr),
\label{lind3}
\end{eqnarray}
where $\gamma_{xj}$ are corresponding loss rates and the Lindblad operators are given by 
\begin{equation}
L_{AB,j}=a+y_jb, \quad L_{BC,j}=b+z_jc,   
    \label{ltr}
\end{equation}
where $y_j$, $z_j$ are the asymmetry parameters. We consider only the propagation from the part 1 to the part 2. 

To uncover specifics of the non-reciprocal circulation between modes A, B and C performed by the device shown in  Fig.~\ref{fig4}(a), let us consider first the asymptotic regime when the single-photon uncorrelated loss is absent and  the lengths of both parts are sufficiently large for transition to the stationary  state. From Eqs. (\ref{lind3})
one gets the following relation between the input and output values of the modal amplitudes
\begin{equation}
\begin{bmatrix} \langle a_{out} \rangle \\ \langle b_{out} \rangle \\ \langle c_{out} \rangle \end{bmatrix} = \mathbf{S}_j \begin{bmatrix} \langle a_{in} \rangle \\ \langle b_{in} \rangle \\ \langle c_{in} \rangle \end{bmatrix},
\quad
 \mathbf{S}_j= \frac{1}{F_j}\begin{bmatrix} (y_jz_j)^2  & -y_jz_j^2  & -y_jz_j   \\
 -y_jz_j^2  &  z_j^2 & -z_j \\ 
 -y_jz_j &  -z_j  &  1 \end{bmatrix} 
\label{srot0}
\end{equation}
where $F_j=z_j^2(1+y_j^2)+1$. From Eq.(\ref{srot0}) one has
\begin{equation}
 \left|\frac{[\mathbf{S}_1\mathbf{S}_2]_{12}}{[\mathbf{S}_1\mathbf{S}_2]_{21}} \right|=\left|\frac{y_1}{y_2}\right|, \quad \left|\frac{[\mathbf{S}_1\mathbf{S}_2]_{23}}{[\mathbf{S}_1\mathbf{S}_2]_{32}} \right|=\left|\frac{z_1}{z_2}\right|, \quad \left|\frac{[\mathbf{S}_1\mathbf{S}_2]_{13}}{[\mathbf{S}_1\mathbf{S}_2]_{31}} \right|=\left|\frac{y_1z_1}{y_2z_2}\right|.
\label{srot00}
\end{equation}
So, whereas it is hardly possibly to realize an ideal circulator (which provides for the unidirectional transfer, i.e.  $A\rightarrow B$, $B\rightarrow C$ and $C\rightarrow A$) with the device Fig.~\ref{fig4}(a), nevertheless, energy transfer between the waveguides is strongly asymmetric. 

In Fig.~\ref{fig4} (b-d) we show an example of asymmetric optical circulation for the bipartite structure as given by Eqs. (\ref{lind3},\ref{ltr}) with the all equal  decay rates $\gamma_{x,j}=\gamma$, $\forall x,j $, and mirrored asymmetry parameters in different parts, $y_1=-z_2=-0.9$, $y_2=-z_1=-0.1$. In all panels Fig.~\ref{fig4} (b-d) solid, dashed and dash-dotted lines correspond to the modal amplitudes of the modes a, b, c. The panels  (b,c,d) correspond to the initial coherent excitation with the unit amplitude of the modes a,b,c with vacuum of other two modes.  The transformation of the modal amplitudes at the end of the part 2 as shown in Fig.~\ref{fig4} (b-d) is described by the following scattering matrix
\begin{equation}
\mathbf{S}_1\mathbf{S}_2= \begin{bmatrix} 0.0075  & 0.0070  & -0.0529   \\
 0.0692  &  0.0669 & -0.5306 \\ 
 -0.0768 &  -0.0743  &  0.5897 \end{bmatrix}. 
\label{srot}
\end{equation}
The matrix (\ref{srot}) shows of about 20 dB asymmetry in power transfer between the waveguides A and B, and B and C. 
 Notice that due to strong coupling asymmetry even for the interaction time much exceeding the inverse decay rate $\gamma^{-1}$, the dynamics shown in Fig.~\ref{fig4} (b-d) is still not close to the asymptotic regime. 

Dynamics shown in Fig.~\ref{fig4} (b-d) demonstrate another curious  feature: namely, that the asymmetry in state distribution does not always require and is not always connected to breaking of reciprocity (for the discussion, see, for example, Ref. \cite{PhysRevApplied.10.047001})   Each part of the structure depicted in Fig.~\ref{fig4}(a)  does break reciprocity by itself. However, for example,  the  part 1 by itself provides for quite asymmetric state propagation from the waveguide A to the waveguide B, and from the waveguide C to the waveguide B. Excitation of the waveguide A creates strong excitation of the mode B, whereas excitation   the mode C leaves the mode B non-exited.  It is interesting to notice that an asymmetric but reciprocal state distributor was suggested in Ref.\cite{Mukherjee2017DissipativelyCW} for a dissipatively coupled chain of waveguides. There excitation was shown to spread either to the left waveguide chain or to the right one in dependence on the excitation of the control waveguide.

\section{Conclusions}

Here we have shown that dissipatively coupled systems can break reciprocity even without simultaneous conjoining unitary and dissipative coupling. Even for the linear systems, dynamics corresponding to different dissipative coupling is not commutative. This can give rise to practical linear optical diode devices providing for unidirectional propagation of certain classes of states. We outlines the way to fabricate such devices using laser-written waveguides in glass or generic integration platform on the basis of such standard materials as InP (in the Appendix we outline practical ways to create such planar integrated structures). 

We show also that designed nonlinear loss enables one to create a single-mode optical insulator. We demonstrate the way to do that with nonlinear coherent loss, and show it is possible to achieve a single-mode non-asymptotic breaking of reciprocity with a practically realizable system of waveguides with self-Kerr nonlinearity. Such system can realize both two-photon loss and nonlinear coherent loss able asymptotically to produce a single-photon state from the initial coherent state. We also show that dephasing is able to induce non-reciprocity, and demonstrate non-reciprocity of correlation transfer with a simple example of bipartite system including a beam-splitter and a dephaser. We demonstrate  an example of non-reciprocal optical circulator is a three-mode system with purely dissipative coupling.

\begin{figure}[t!]
\centering
\includegraphics[width=0.8\linewidth]{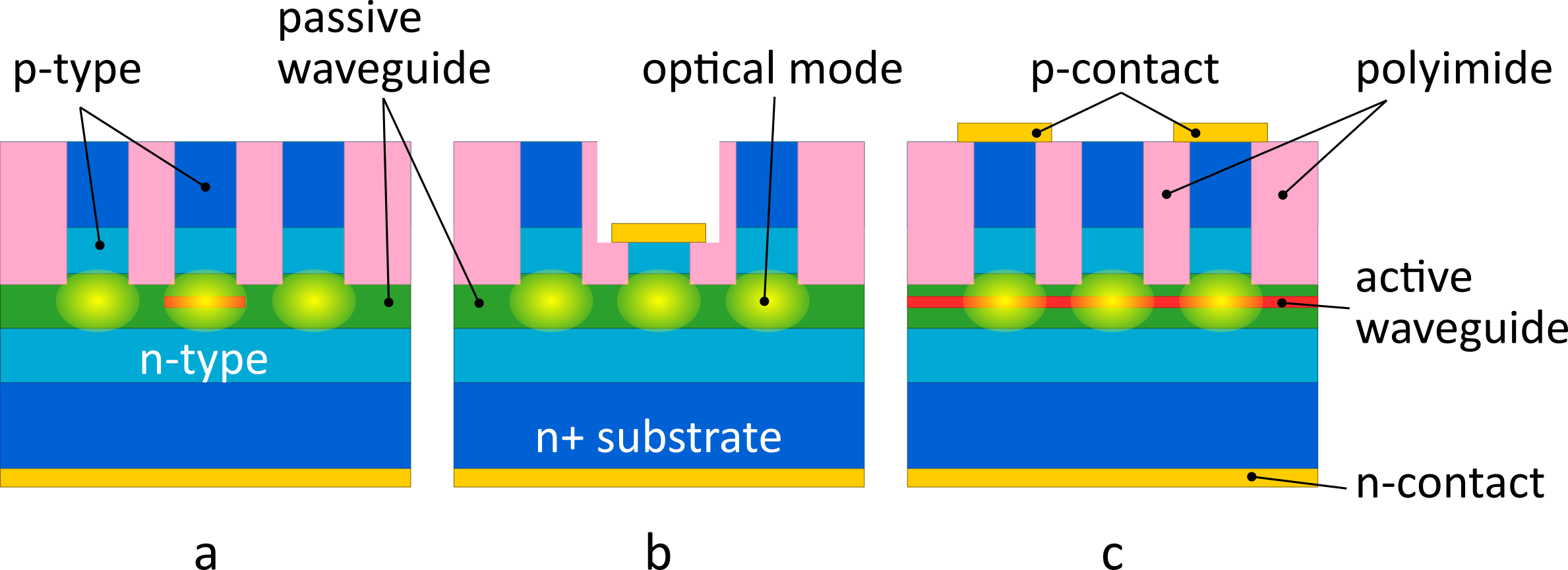}
\caption{Possible implementations of the dissipative coupler. a -- active unbiased waveguide, b -- passive waveguide in isolation cross-section with metal, c -- active waveguides with applied bias. In each cross-section, two side waveguides and a middle waveguide with extra loss are shown.}
\label{fig:implementation_options}
\end{figure}

\section*{Acknowledgments}

The authors (D.M. and I.P.) gratefully acknowledges support from the EU project PhoG 820365 and the NATO project NATO SPS - G5860; D.P. would like to acknowledge the Dutch Stimulus OPZuid programme through the Open Innovation Photonic ICs project (PROJ-00315). The authors thank Prof. K. Williams for useful criticism and fruitful discussion. 

\section*{Appendix: Possibilities for practical implementation}

In this section we propose several options for practical realization of dissipative coupling which can be used to construct the schemes discussed before (i.e., constitute components of the bipartite set-ups discussed there). The suggested solutions use a mature COBRA / Smart Photonics InP integration platform \cite{jstqe-augu18} which features a large set of standard building blocks, and thus can be used to integrate the dissipative couplers in more complex circuits. 

The suggested dissipative coupler consists of three parallel single-mode waveguides A, B, and C, where C corresponds to the common loss reservoir as shown in Fig.\,\ref{fig1}(a). The waveguide C is placed between A and B, and has excessive loss $\Gamma$ as compared to them. The light coupling occurs in the waveguide pairs A--C, and C--B. % The three options for obtaining an additional controlled loss are discussed below.

\textbf{Platform description.} In the selected integration platform \cite{jstqe-augu18}, there are two types of waveguides: active and passive, see Fig.\,\ref{fig:implementation_options}. The \textit{passive} refers to a lattice-matched InGaAsP material which is transparent in the C band (1530--1565\,nm). It is used as a waveguiding layer with InP as upper and lower cladding layers with a lower refractive index. The \textit{active} is a quantum well-based structure (QW), which uses quaternary InGaAsP material with a composition chosen to have the emission wavelength and bandgap around 1550\,nm. This leads to presence of absorption at this wavelength and provides means to create controlled amount of excess loss in waveguide C.

We have considered three possible ways of achieving excess loss in the middle waveguide, shown in Fig.\,\ref{fig:implementation_options}.

\textbf{Unbiased active sections.} This option is shown in Fig.~\ref{fig:implementation_options}a. The middle waveguide is an active waveguide with a high absorption at 1550\,nm if unbiased \cite{pj-pust18}. This is the most natural solution considering the absorption spectra of the active and passive waveguides. However, taking into account the required distance between the waveguides (below 3\,$\mathrm{\mu m}$), the reliable integration of such narrow active stripes by passive regrowth is not possible with the current fabrication flow.

\textbf{Passive waveguides with metal.} This option is shown in Fig.\,\ref{fig:implementation_options}b. The middle waveguide is fabricated in the isolation cross-section, in which the strongly conductive part of the top p-type cladding layer is removed. If we add the metal on top of this waveguide, it will overlap with the optical mode and cause extra losses. This option does not require any electrical signals to be applied to the coupler. However, there is no possibility to tune the excess loss on the middle waveguide.

\textbf{Active waveguides with applied bias.} This option is shown in Fig.\,\ref{fig:implementation_options}c. All three waveguides are fabricated in the active cross-section, and the two side waveguides have metal contacts connected on top of them. By applying reverse bias voltage or pump current to these metal pads, one can control the absorption in the QW layer, and thus tune the amount of excess loss in the middle waveguide. This scheme allows some tuneability, but requires active driving electronic to be permanently connected. And similar to the first option, small distance between the waveguides means that the opening between the metal stripes will be 3 to 4~$\mathrm{\mu m}$, which is on the limit of the fabrication technology.

%\bibliography{bibl}

%apsrev4-2.bst 2019-01-14 (MD) hand-edited version of apsrev4-1.bst
%Control: key (0)
%Control: author (8) initials jnrlst
%Control: editor formatted (1) identically to author
%Control: production of article title (0) allowed
%Control: page (0) single
%Control: year (1) truncated
%Control: production of eprint (0) enabled
%

\end{document}